Consideration of the Need for Quantum Grid Computing

Type: *Short Research Paper*

Dominic Rosch-Grace
Department of Computer Science
North Dakota State University
1320 Albrecht Blvd., Room 258
Fargo, ND 58102
dominic.rosch@ndsu.edu

Jeremy Straub            *** *Corresponding Author* ***
Department of Computer Science
North Dakota State University
1320 Albrecht Blvd., Room 258
Fargo, ND 58102
jeremy.straub@ndsu.edu

**Abstract**

Quantum computing is poised to dramatically change the computational landscape, worldwide.  Quantum computers can solve complex problems that are, at least in some cases, beyond the ability of even advanced future classical-style computers.  In addition to being able to solve these classical computer-unsolvable problems, quantum computers have demonstrated a capability to solve some problems (such as prime factoring) much more efficiently than classical computing.  This will create problems for encryption techniques, which depend on the difficulty of factoring for their security.  Security, scientific, and other applications will require access to quantum computing resources to access their unique capabilities, speed and economic (aggregate computing time cost) benefits.  Many scientific applications, as well as numerous other ones, use grid computing to provide benefits such as scalability and resource access.  As these applications may benefit from quantum capabilities – and some future applications may require quantum capabilities – identifying how to integrate quantum computing systems into grid computing environments is critical.  This paper discusses the benefits of grid-connected quantum computers and what is required to achieve this.

**1. Introduction**

As humanity produces ever more data and develops new ways of processing and reasons to process it, the demand for information processing capability, efficiency, and the accessibility of computing resources necessarily increases. Grid computing is a response to the need to effectively use computing resources for large and processing intensive applications.  A grid computing job can, conceivably, use computers from all over the world, which are part of a shared computing system, minimally constrained by geographic considerations.  However, types of computing problems exist that cannot be efficiently solved by marshalling hordes of classical computing resources.

Computing problems can be placed into several classes.  The class 'P' includes problems that classical computers can efficiently solve.  The class 'NP' includes problems that classical computers cannot solve efficiently but can efficiently verify the correctness of an answer for.  Class P problems are effectively solved by classical computers. Class NP problems (such as the prime factoring used in modern cryptography [1]) can benefit – potentially significantly – from quantum capabilities, but are solvable by classical computers [2].  The classes 'P' and 'NP' both are subsets of a larger class called 'PH', which are classical computer solvable problems.  Another class, called 'BQP' includes problems that are efficiently solved by quantum computers.  Recently, Raz and Tal demonstrated that there was a problem within the class BQP which was not within PH.  This means, as Hartnett [3] termed it, that it is "a problem that only quantum computers will ever be able to solve".  That is that, irrespective of future advances – absent of a fundamental architectural level change, the problem must be solved on quantum computers because it cannot be solved on classical ones.

Quantum computers are unconstrained by the binary mechanics present in traditional computing devices, resulting in increased processing capability for many classically solvable problems. They also have the inherent capability to handle uncertainty and probabilistic calculations. As demonstrated by Shor's Algorithm [4], the impact of the differences between quantum and classical processing can be significant. Thus, while grid computing seeks to maximize the benefits of a networked collaboration of devices [5], it cannot – unaided – rival quantum performance for certain problems. As quantum computing's utility has been demonstrated for some machine learning techniques [6] and in key grid-use disciplines such as chemistry [7] and materials science [8], among others, grid systems without quantum capabilities are not able to support the full range of potential user needs.

This paper asks and seeks to answer two questions: what are the benefits of grid-connected quantum computers and what is needed to connect quantum computers to the grid? It continues, in Section 2, with a discussion of grid computing benefits. Section 3 discusses the benefits of quantum computing. Section 4 reviews existing cloud computing quantum capabilities, while Section 5 discusses what is needed to connect quantum computers to grid systems. Section 6 considers the benefits of grid connected quantum computers, before the paper concludes and discusses necessary future work in Section 7.

## 2. Grid Computing and Its Benefits

Grid computing is a virtual, decentralized collaboration of potentially heterogenous machines over the internet to optimize workload and share information in pursuit of mutual, large-scale computational goals. The grid computing paradigm is analogous to an electrical grid, in the sense that users have access to computational resources with minimal consideration as to where the resources come from [9]–[11]. The principal arguments for grid computing are based on empirical evidence that computing resources (such as supercomputers and computer clusters) are expensive to procure, operate, maintain, and utilize. Thus, the grid paradigm serves as a means of making computing capability accessible, regardless of users' geographical location [12]. A key benefit of the grid infrastructure is the exploitation of otherwise under-utilized resources. Under-utilization of computational resources tends to be common in modern business practices; this applies to devices such as desktop computing equipment, and disk drive capacity, for example [13].

## 3. Quantum Computing and its Benefits

Quantum computing is a powerful new class of processing hardware that harnesses quantum mechanical principles of superposition and entanglement. Currently, quantum computers have limited capabilities; however, these capabilities are consistently growing. The elementary unit of information in quantum computers is a qubit. Qubits are the quantum alternative to bits, in the sense that characteristics such as their quantum state (electron spin or photon polarization) represents information on the theoretically infinitely subdivisible range from 0 to 1. This quantum mechanical phenomena is referred to as superposition [14]. This provides the capability to generate truly random numbers, a characteristic in which traditional computation can only emulate [15]. Conversely, typical bits operate on a binary mechanic, where they are limited to either 0 or 1 [16], [17]. The celerity of advancement in quantum processing capability is denoted by metrics such as the number of qubits and attenuated error rates for a selection of logical operations. In 2018, Google's Quantum AI lab presented "Bristlecone," a quantum processor comprised of a two-dimensional array of 72 qubits. The development of such a processor, coupled with the statistics regarding diminished error rates associated with including more qubits is a product of the drive in industry for improved quantum processing capabilities [18]. Although empirical evidence suggests there is a mathematical relationship between number of qubits and mitigated error rates derived from quantum logical correction operations, simply increasing machine size is not an immediate solution to reducing error given the difficulty of qubit management.

Computational complexity theory indicates that certain mathematical endeavors tasked to parallel quantum circuits are independent of their intrinsic size; that is, quantum advantage means that the steps required to complete a given task operates on a mathematically logarithmic basis [19]. This characteristic of quantum computational systems does not occur on classical circuits. Beyond comparison, there is no generalized explicit means of identifying whether a task benefits from an inherent quantum advantage. This area remains a topic of discussion, as there is no recognized metric for assessment [19]. Presently, the demonstration of quantum advantage comes from implementation of mathematically complex tasks on both quantum and traditional systems and comparing the results. For some applications, the information processing potential of quantum computers is apparent. Shor's implementation of a quantum integer factorization algorithm, in 1994, exemplified the disruptive capabilities of quantum systems and increased interest in quantum computing research within the scientific community [20]. Shor's Algorithm was created as a demonstration of how quantum systems can render classical cryptography obsolete by effectively identifying the prime factors of a large integer. [21] The demonstration of quantum advantage, in conjunction with the applications for quantum systems, has generated significant interest in computational research in quantum computing.

**4. Beyond Quantum Cloud Computing**

Cloud computing provides users with computational resources over the internet [22]. The implementation of a quantum-cloud interface gives users the ability to access quantum computational resources, irrespective of their geographical location. This may be utilized for a variety of applications, including security. Quantum cryptography may be a key initial application for quantum computing as it provides significantly improved security for sensitive data stored or moving within the cloud [23]–[25]. Enhanced security may increase cloud versatility and enhance trust among existing and prospective users [26]–[28], potentially leading to greater cloud computing adoption and use.

Prior work in quantum cloud computing provides several frameworks for quantum encryption and data management. Qazi and Aizenberg [23] advocate on behalf of a quantum algorithm that demonstrates successful management of data in the cloud, exemplified by proof-of-concept studies on applicable load traces. This algorithm was compared to modern, established host-load algorithms to demonstrate accuracy [23]. As discussed previously, quantum advantage applies to mathematically complex endeavors that classical computing cannot practically achieve. A cloud-based quantum infrastructure could be used to provide quantum services over a network to users worldwide.

Despite the infancy of quantum computing, various technological innovators have provided a means of online quantum exploration and use. In 2017, IBM launched "IBM Q," a quantum service based off of their 2016 service, "Quantum Experience." Here, users have access to quantum computational resources over the internet and consequently can delve into the research and application development necessary to produce key quantum algorithms [29]. Likewise, Microsoft offers "Azure Quantum," a public, cloud-based ecosystem providing accessible means of 'full-stack' quantum software development. Microsoft quantum services utilize the Quantum Development Kit (QDK), a programming language exclusive to quantum resources (Q#), as well as a mutual, open-source interface of quantum platforms and a selection of languages [30].

**5. Bringing Quantum Capabilities to the Grid**

Cloud-based quantum computing can facilitate the expeditious availability of resources and thus enable scientific progression in fields such as physics, chemistry, cryptography, and machine learning. Further, the availability of quantum services via the cloud eliminates many geographical barriers to access. However, while cloud-based quantum computing provides a multitude of benefits, it is still limited by several key logistical challenges, as cloud-based services do not necessarily provide the resource access, scheduling, and job management capabilities essential for many computational processing applications.

A quantum grid architecture would provide the workload optimization capabilities present in traditional heterogeneous grid environments, as well as the potential for similar efficiency for mathematically complex problems allocated to a quantum computer. Existing uses of quantum capability, such as Shor's algorithm, demonstrate that large integer factorization is done significantly faster on quantum computing; this drives the clear need for quantum resource access for cryptographic frameworks. However, this is only one of several applications areas that would benefit from quantum grid computing.

To provide the benefits of a grid-based computational framework, it is important to optimize the distribution of tasks across the various nodes. The distribution and allocation of tasks, in accordance with what computational resources are available to the grid network, is referred to as grid scheduling. Prior heterogeneous grid scheduling algorithms can prospectively be modified to support quantum hardware in the grid. Such an algorithm would necessarily consider if assignment of tasks to a quantum computer would result in a quantum advantage.

A heterogenous grid is comprised of a multitude of devices, all of which utilize middleware for collaboration among devices. The format of the middleware would need to be compatible with both quantum and traditional systems to allow collaboration between all nodes in a grid network, including quantum hardware. Given this, the availability of middleware providing quantum-compatible functionality for all devices within the network is required for the successful development of a quantum grid infrastructure.

In addition to middleware, the development of a quantum agnostic intermediate language (IL) could provide significant flexibility for scheduling jobs between quantum and classical computers. By storing the jobs in this IL, jobs could be split between the two processor types on the fly, by the scheduler.  In some cases, even jobs with a notable quantum advantage may be assigned to classical computers, if dictated by availability.  Notably, Microsoft has demonstrated this conceptually with the development of their quantum computer emulator, as part of their Quantum Development Kit.  This emulator allows Q# code to run on classical computers, in addition to quantum ones.

While the IL would not be necessary for a first-generation quantum computing grid, it would be an enhancement that would enable scheduling flexibility.  The IL could also have a key role in supporting multiple types of quantum processing systems, which may have different underlying formats and command structures, which could be supported through the compilation-to-machine-code process.

**6. Quantum Capabilities and their Benefits on the Grid**

The computational advantage of a quantum computer, paired with the efficiency of grid architecture is poised to provide benefits to a spectrum of computing applications, if an optimal allocation of tasks can be achieved. Removing tasks that can be more efficiently performed by quantum computers from consuming extensive resources on classical computers frees up these resources for other jobs that can be efficiently performed on classical computers. It is not necessary to decide this at the program level. For example, certain tasks within a program, such as producing a random number, can be allocated by the control system to a quantum computer, which has the capability to generate true random numbers [31]. Meanwhile, other components of the program or system can be left to operate on classical devices.

In other cases, entire programs – such as those with complex mathematics – that are projected to take an impractical amount of time running on a traditional computer, can be assigned exclusively to a quantum computer. The quantum computer can return results to the command system to distribute to the user or to transfer to other computers for more processing.

This processing power and versatility can provide computational capabilities to groundbreaking research on a computing cost basis, instead of requiring system procurement and operational costs to be covered by a single party. Quantum computing availability may unlock new scientific computing and artificial intelligence capabilities for a variety of fields. An important aspect of scientific computing and artificial intelligence is the ability to process and apply information [32].  Among other applications, the information processing potential in a quantum grid may support the development of a more generalized artificial intelligence [5], [19], [33]–[36]. Such a capability could help to solve many of the world's most difficult problems, which humans are incapable of processing given the tools at their disposal, in conjunction with biological limitations. Another application likely to result from quantum grid computing is improved cyber defense, as modern encryption is dependent on mathematical frameworks that prevent sensitive data from being obtained by a third party. The computational ability of the quantum grid allows for an enhanced degree of flexibility and complexity in the construction of essential cryptographic frameworks to replace those that will be rendered obsolete by widely available quantum computing [36]–[38].

**7. Conclusions and Future Work**

This paper has considered the need for quantum grid computing and discussed what steps are required to achieve this goal.  From the foregoing sections, it is clear that there is significant potential value in integrating quantum computing capabilities into the grid architecture and using grid infrastructure to facilitate access to quantum computing capabilities.  Benefits include the ability to handle certain types of algorithms that cannot be solved (or cannot be efficiently solved) on classical computers as well as the potential to intelligently split computing jobs between classical and quantum computers, depending on the exact capabilities requested or required by the application.

In order to make these benefits possible, additional technological development is required. A number of key logistical challenges have been identified. These include the need for a technical capability to automatically ascertain what level of benefit a piece of code would receive from running on a quantum computer, for prioritization purposes, and developing a system for scheduling jobs between the two heterogeneous systems. To support this scheduling flexibility, code would need to be implemented in a such a manner that will run (excepting quantum-solvable-only problems) on both types of platforms so that it can be automatically tasked to the platform type and particular machine identified by the scheduler.

With large-scale quantum grid computing, there is the potential for significant data analysis-driven societal gains. The quantum grid can shepherd computational capabilities to the most unknown and groundbreaking fields. While quantum-enabled grid capabilities may hasten the obsolescence of certain forms of cryptography, they may also bring a new era of faster and more powerful artificial intelligence and help to answer questions across numerous fields of academic inquiry.

**Acknowledgements**

This work has been partially supported by the U.S. National Science Foundation (NSF award # 1757659).